\documentclass[letterpaper,twocolumn,prl,showpacs,groupedaddress]{revtex4-1}

\usepackage[draft]{hyperref}
\usepackage{amsmath}
\usepackage[pdftex]{graphicx}   %for the pdftex use
\usepackage{dcolumn}
\usepackage{amssymb,amsmath}
\usepackage{color}
\usepackage{natbib}

\newcommand{\ba}{\begin{array}}
\newcommand{\ea}{\end{array}}
\newcommand{\be}{\begin{equation}}
\newcommand{\ee}{\end{equation}}
\newcommand{\bea}{\begin{eqnarray}}
\newcommand{\eea}{\end{eqnarray}}
\newcommand{\bnabla}{\boldsymbol{\nabla}}

\begin{document}

%%%%\twocolumn[ %% activate for two-column option

%\title{Title}
\title{Universal morphologies of fluid interfaces deformed by the radiation pressure\\of acoustic or electromagnetic waves}

%\author{Authors}
\author{N. Bertin, H. Chra\"ibi, R. Wunenburger*, J.-P. Delville, and E. Brasselet*}

\affiliation{Univ. Bordeaux, LOMA, UMR 5798, F-33400 Talence, France.\\
CNRS, LOMA, UMR 5798, F-33400 Talence, France.}

\date{\today}

\begin{abstract}
We unveil the generation of universal morphologies of fluid interfaces by radiation pressure whatever is the nature of the wave, acoustic or optical. Experimental observations reveal interface deformations endowed with step-like features that are shown to result from the interplay between the wave propagation and the shape of the interface. The results are supported by numerical simulations and a quantitative interpretation based on the waveguiding properties of the field is provided.
\end{abstract}

\pacs{43.25.Qp, 42.50.Wk.}

% 43.25.Qp Radiation pressure [in Nonlinear acoustics]
% 42.50.Wk Mechanical effects of light on material media, microstructures and particles

%%%%] %% activate for two-column option

\maketitle

More than one century after the pioneering works of Poynting~\cite{loudon_12} and Rayleigh~\cite{beyer_78} on the mechanical effects of electromagnetic and acoustic waves, it is a common knowledge that both light and sound exert radiation pressure on matter. Experimentally, a striking demonstration of its existence is the observation of deformations of fluid interfaces as shown in acoustics by Hertz and Mende in 1939~\cite{hertz_39} and in optics by Ashkin and Dziedzic in 1973~\cite{ashkin_73}. Since these observations, radiation pressure of waves have been exploited in various contexts such as contactless metrology of fluids (\cite{ostrovskaya_88} in optics and~\cite{khuri_88} in acoustics),  liquid droplets ejection (\cite{zhang_88} in optics and \cite{elrod_89} in acoustics), and biomedical applications (\cite{guck_00} in optics and~\cite{fatemi_98} in acoustics).

Here we report on universal kind of steady state deformations of initially flat fluid interfaces induced by radiation pressure whatever is the nature of the wave, be it acoustic or optical. We show that this novel family of step-like universal interface morphologies results from the balance between buoyancy, capillarity and radiation pressure for both the acoustic and electromagnetic cases. Performing numerical simulations we show that refraction drives the interplay between the shape of a deformed interface and the wave propagation along it. Finally, we propose an interpretation of the observed morphologies based on the waveguiding properties of the deformations.

%---------------------------------------------------------------------------------------------------
\begin{figure}[b]
\centering\includegraphics[width=0.85\columnwidth]{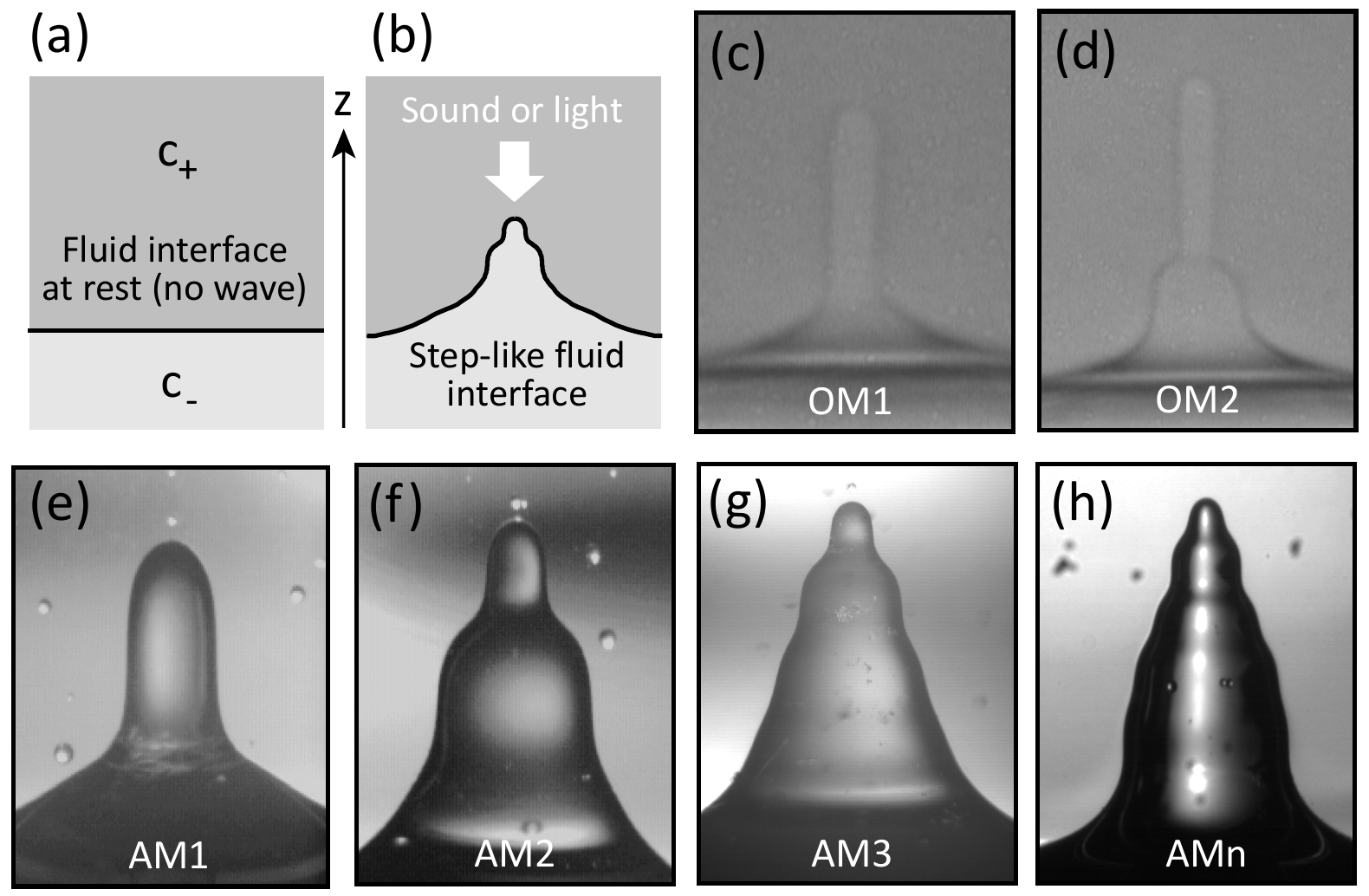}
\caption{\label{fig1}
(a) Planar fluid interface at rest. (b) Step-like interface deformed by radiation pressure. (c,d) First (OM1) and second (OM2) order optical morphologies, which correspond to input beam power 470mW and 1020mW, respectively. (e$-$h) First (AM1), second (AM2), third (AM3) and higher order (AM$n$) acoustic morphologies, which correspond to input beam power 4.0W, 1.0W, 220mW and 20mW, respectively.}

\end{figure}
%---------------------------------------------------------------------------------------------------

% AM1 : param4     % AM2: param3    % AM3 : param1  % AMn : param2
% cf. Fig.A1 these Bertin
%---------------------------------------------------------------------------------------------------
\begin{table*}
\begin{ruledtabular}
\begin{tabular}{ccccc}
   & AM1 & AM2 & AM3 & AM$n$ \\
\hline
Fluid ``$+$''                                        & Water      & Water                         & Salted water 25wt.\% & Silicone oil 100cSt\\
Fluid ``$-$''                                         & Kerosene & Silicone oil 100cSt   & Chloroform                    & Oil FC72\\
$c_+$ (m $\cdot$ s$^{-1}$)             & 1490 &        1490                          & 1783                                & 1000 \\
$c_-$  (m $\cdot$ s$^{-1}$)             & 1315 &        1000                           & 1000                               & 512 \\
$\rho_+$ (kg $\cdot$ m$^{-3}$)     & 998    &         998                           &  1189                               & 966 \\
$\rho_-$ (kg $\cdot$ m$^{-3}$)     &  790   &         966                            &  1500                               &  1680 \\
\end{tabular}
\end{ruledtabular}
\caption{\label{table1}
Fluid characteristics at room temperature for the acoustic experiments
%that lead to morphologies labeled AM1, AM2, AM3 and AM$n$, see Fig.~\ref{fig1}
, where $c_\pm$ and $\rho_\pm$ refer to the phase velocity and the density of the fluids labeled ``+" and ``$-$", respectively.}
\end{table*}
%---------------------------------------------------------------------------------------------------

This is made possible by using interfaces between simple fluids for acoustic waves whereas extremely soft interfaces of near critical fluids are used in the case of optical waves. At rest, two immiscible fluids define a planar interface, see Fig.~\ref{fig1}(a). The interface is deformed by the acoustic or optical radiation pressure of a focused sound or light beam that impinges at normal incidence from the medium with the largest phase velocity ($c_+$) to the one with the lowest phase velocity ($c_-$), thereby leading to step-like deformations as sketched in Fig.~\ref{fig1}(b).

Experimentally, the acoustic beam is obtained from an immersed, spherical, piezoelectric transducer with a 38.4~mm radius of curvature, 38~mm diameter operating at 2.25~MHz central frequency. In the optical case, we use a focused Gaussian beam at 514.5~nm wavelength. For each experiment, we set the incident power to a value that gives an aspect ratio $h/w_0$ of a few units, where $h$ is the height of the fluid interface deformation and $w_0$ is the beam waist~\cite{remark_waist}. Typical examples of the observed morphologies are displayed in Fig.~\ref{fig1}(c)$-$\ref{fig1}(h). In addition, the relevant characteristics of the fluids used in the acoustic case are summarized in Table~\ref{table1} whereas in optics we use the transparent, isotropic and nonmagnetic two-phase microemulsion described in~\cite{casner_01}. At $T-T_c=4~K$, where $T_c$ is the critical temperature above which two distinct phases coexist, $c_+ = 0.6855 c$ and $c_- = 0.6807 c$ where $c$ is the speed of light in vacuum.

Hereafter, the observed fluid interface morphologies are labeled following their number $n$ of steps, namely AM$n$ and OM$n$ in the acoustic and optical case, respectively. Quantitatively, we demonstrate that the steady-state morphologies obey the following equilibrium equation between radiation and gravito-capillary pressures,
\be
\label{eq_equilibre}
\Pi_{\rm R} = \Pi_{\rm GC}\,,
\ee
In Eq.~(\ref{eq_equilibre}) the radiation pressure is $\Pi_{\rm R} = {\bf n} \cdot \left[  \left(\langle{\mathbb T}_+\rangle - \langle{\mathbb T}_-\rangle \right)  \cdot {\bf n} \right]$ with ${\bf n}$ the unit vector normal to the interface oriented from fluid ``$-$" to fluid ``$+$" and $\langle {\mathbb T}_\pm \rangle$ refers to the time averaged values of the radiation tensor at both sides of the interface over a wave cycle. The gravito-capillary contribution $\Pi_{\rm GC} = |\rho_+ - \rho_-|g h - \frac{\gamma}{r}\frac{d}{dr}[rh'(r)/\sqrt{1+h'^2(r)}]$~\cite{delville_05} where $\rho_\pm$ is the density of fluid $\pm$, $g$ is the modulus of the gravitational acceleration, $\gamma$ is the interfacial tension, and $h(r)$ is the height of the axisymmetric deformed interface, with $h=0$ at rest and $h>0$ when the wave is turned on, and $h'(r) = \frac{dh}{dr}$. In addition, we have introduced the cylindrical coordinate system $(r, \phi, z)$ with orthonormal basis $({\bf e}_r, {\bf e}_\phi, {\bf e}_z)$, $z$ being the symmetry axis of the deformation oriented toward its tip (see Fig.~\ref{fig1}).

On the one hand $\Pi_{\rm GC}$ is calculated from the experimental interface profile $h(r)$. On the other hand, the evaluation of $\Pi_{\rm R}$ requires the computation of the field. This is done by solving the propagation equation for a given experimental profile $h(r)$ accounting for the continuity relations for the field at the interface and appropriate source distribution that defines the incident beam.

\begin{figure*}[t]
\centering\includegraphics[width=1.9\columnwidth]{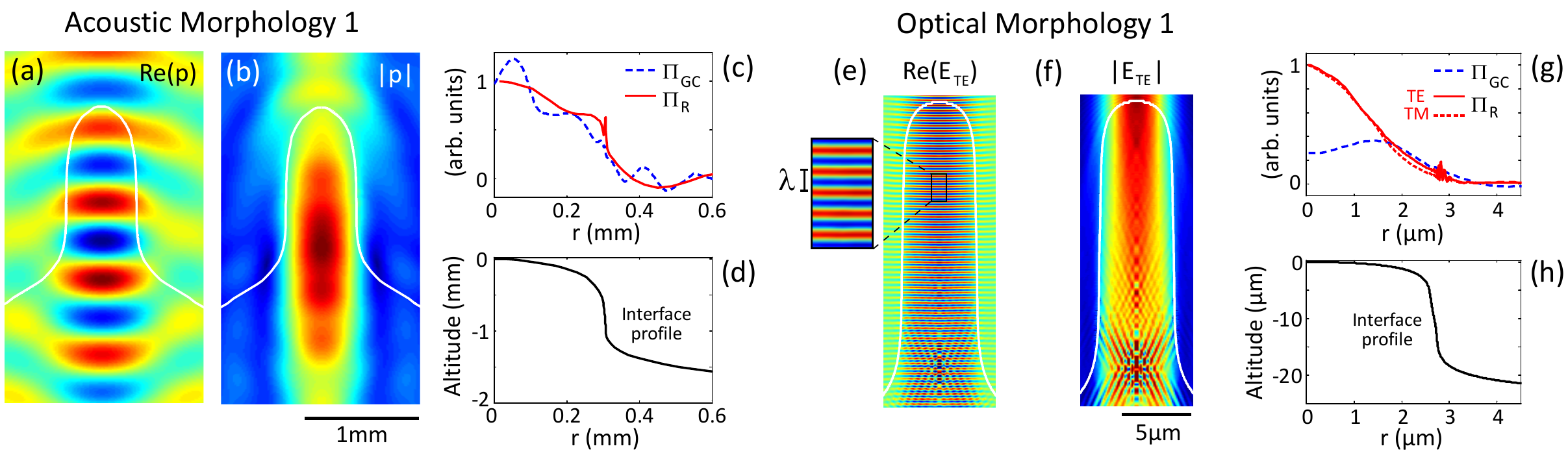}
\caption{\label{fig2}
(color online) Analysis of the AM1 and OM1 cases, see Fig.~\ref{fig1}(e) and Fig.~\ref{fig1}(c). In each case, the spatial distribution of the real part and the modulus of the field in an equatorial plane are shown with the deformed interface profile superimposed on it as a white solid curve [panels (a,b) and (e,f)]. (c,g) Balance between the radial dependence of the radiation pressure ($\Pi_{\rm R}$) and the gravito-capillary contribution ($\Pi_{\rm GC}$), see~\cite{remark_testEq1} for details. (d,h) Deformed interface profiles. The altitude at the tip of the deformation is set here to zero in all cases.}
\end{figure*}

\begin{figure*}[t]
\centering\includegraphics[width=1.9\columnwidth]{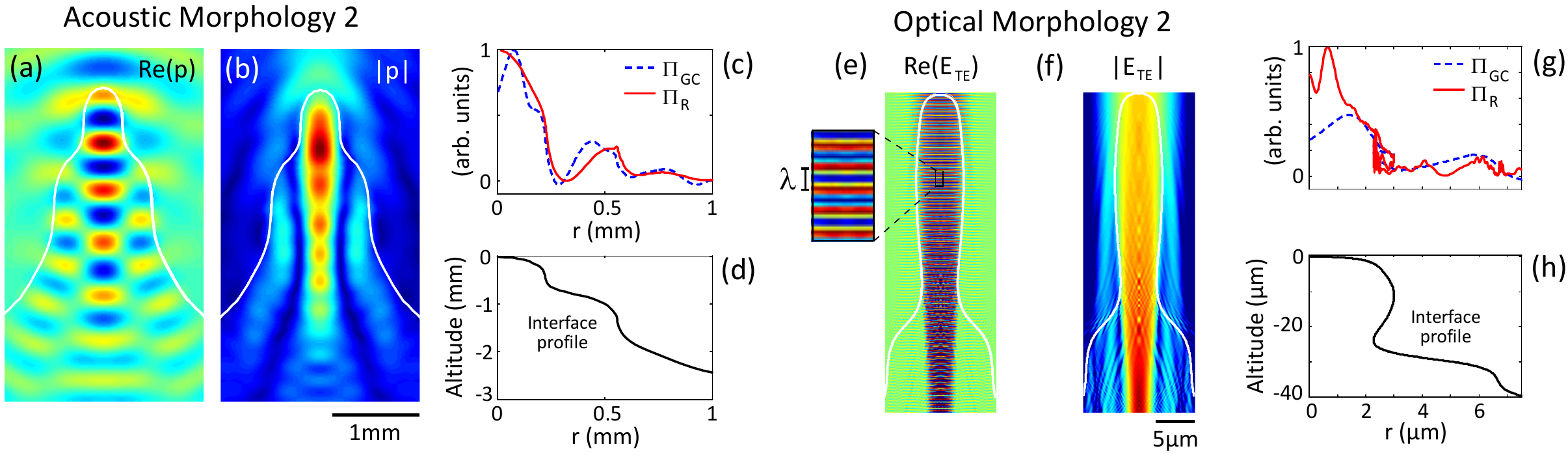}
\caption{\label{fig3}
(color online) As in Fig.~\ref{fig2} for the AM2 and OM2 cases, see Fig.~\ref{fig1}(f) and Fig.~\ref{fig1}(d), and~\cite{remark_testEq1} for details.}
\end{figure*}

\begin{figure}[t]
\centering\includegraphics[width=0.9\columnwidth]{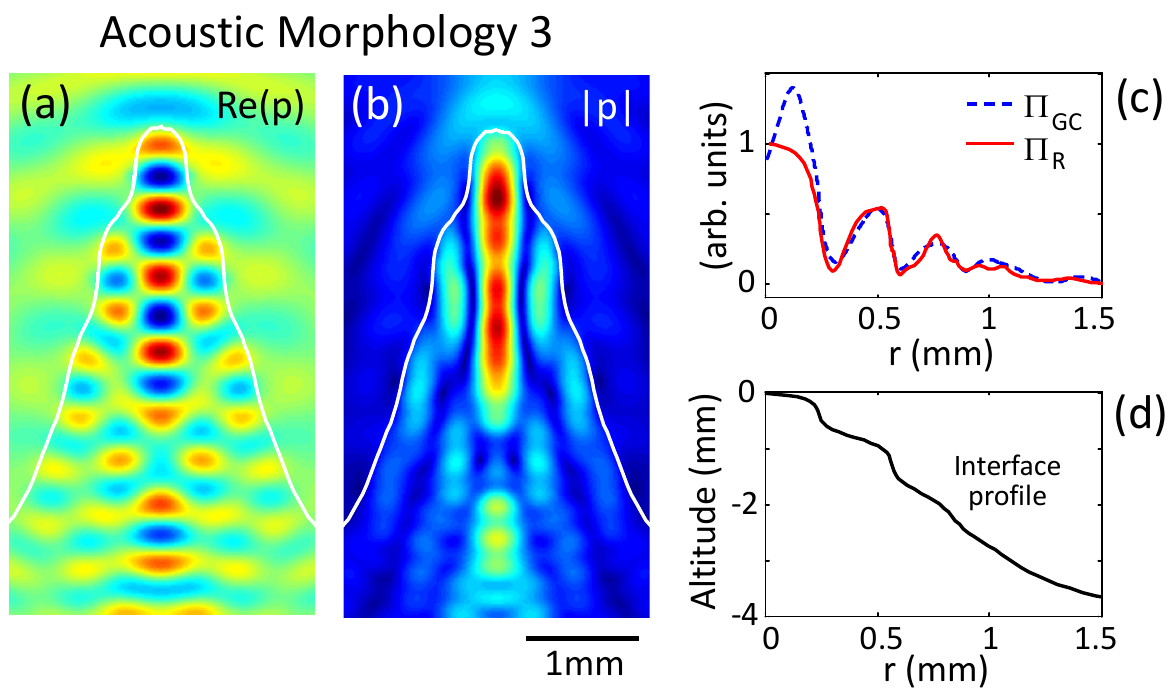}
\caption{\label{fig4}
(color online) As in Fig.~\ref{fig2} for the AM3 case, see Fig.~\ref{fig1}(g), and~\cite{remark_testEq1} for details.}
\end{figure}

In acoustics, the linear propagation equation in perfect fluids expresses as $\Delta \Psi_\pm - \frac{1}{c_\pm^2} \frac{\partial ^2 \Psi_\pm}{\partial t^2} = 0$ where $\Psi_\pm$ is the complex  axisymmetric acoustic pressure field in each fluid, $p_\pm(r,z,t)$, with time $t$. Assuming harmonic fields with angular frequency $\omega$, the acoustic velocity field is obtained from the Euler equation that gives ${\bf u}_\pm(r,z,t) = \frac{1}{i \omega \rho_\pm} \bnabla p_\pm$. Numerically, the propagation problem---Helmholtz equation in a two-phase axisymmetric configuration---is solved by using a boundary element method~\cite{kirkup_98}. Once the pressure and velocity fields are calculated, see panels (a) and (b) in Figs.~\ref{fig2}, \ref{fig3} and \ref{fig4} where the real part and modulus of $p$ are shown for the AM1, AM2 and AM3 situations, $\Pi_{\rm R}$ is computed using Brillouin expression for the acoustic radiation tensor ${\mathbb T}_{\rm ac}$
\vspace{-3mm}
\be
\label{eq_Tac}
{\mathbb T}_{\rm ac} = -\frac{1}{2}\chi p^2 {\mathbb I} + \frac{1}{2}\rho {\bf u}^2{\mathbb I} - \rho {\bf u} \otimes {\bf u}\,,
\ee
where $\chi_\pm=(\rho_\pm c_\pm^2)^{-1}$ is the isentropic compressibility of each fluid, ${\mathbb I}$ is the identity tensor, and $\otimes$ is the dyadic product.
The validity of Eq.~(\ref{eq_equilibre}) is tested by comparing the radial dependence of $\Pi_{\rm R}$ and $\Pi_{\rm GC}$ for various morphologies as shown in panel (c) of Figs.~\ref{fig2}, \ref{fig3} and \ref{fig4}~\cite{remark_testEq1}. We conclude to an overall satisfying validation of Eq.~(\ref{eq_equilibre}). Noticeably, the main extrema of $\Pi_{\rm R}$ correspond to the ones of $\Pi_{\rm GC}$, which sign he step-like nature of the morphologies. We note that the observed on-axis discrepancy between $\Pi_{\rm R}$ and $\Pi_{\rm GC}$ could be ascribed to the hydrodynamic stress exerted at the tip of the deformation by the rectified fluid flow that results from thermoviscous dissipation (``acoustic streaming")~\cite{book_acousticstreaming}.

%---------------------------------------------------------------------------------------------------
\begin{table}
\begin{ruledtabular}
\begin{tabular}{ccc}
 Acoustic wave  & TE optical wave & TM optical wave \\
\hline
$p$             & $E {\bf e}_\phi$  & $H {\bf e}_\phi$\\
${\bf u}$    & ${\bf e}_\phi \times {\bf H}$ & ${\bf e}_\phi \times {\bf E}$\\
$\rho$        & $\mu$ & $\epsilon$ \\
$\chi$        & $\epsilon$ & $\mu$ \\
\end{tabular}
\end{ruledtabular}
\caption{\label{table2}
Table of equivalence between acoustic and electromagnetic wave propagation.
%, where $\epsilon$ and $\mu$ are respectively the dielectric and magnetic susceptibilities assumed to be constant scalars.
}
\end{table}
%---------------------------------------------------------------------------------------------------

In the optical case, since the refractive index contrast is small, $\frac{c_+ - c_-}{c} \lesssim 10^{-2}$, the scalar propagation equation is known to accurately describe the electromagnetic field whatever its polarization state. In order to benefit from the mathematical equivalence between acoustic and electromagnetic wave propagation~\cite{ikelle_12}, we restrict our study to the particular cases of a TE-polarized field, $\Psi = E_{\rm TE}$ where ${\bf E}_{\rm TE} = E_{\rm TE} {\bf e}_\phi$, or TM-polarized field, $\Psi = H_{\rm TM}$ where ${\bf H}_{\rm TM} = H_{\rm TM} {\bf e}_\phi$, where ${\bf E}$ and ${\bf H}$ are the electric and magnetic fields. This allows to use the numerical code developed for the acoustic case following the polarization dependent correspondance summarized in Table~\ref{table2}.
Once the electric and magnetic fields are calculated, see panels (e) and (f) in Figs.~\ref{fig2}, \ref{fig3} and \ref{fig4} that show the real part and modulus of $E_{\rm TE}$ for the OM1 and OM2 situations, $\Pi_{\rm R}$ is computed using Maxwell expression for the electromagnetic radiation tensor ${\mathbb T}_{\rm em}$~\cite{remark_electrostriction}
\be
\label{eq_Tem}
{\mathbb T}_{\rm em} = -\frac{1}{2}\epsilon {\bf E}^2 {\mathbb I} - \frac{1}{2}\mu {\bf H}^2 {\mathbb I} + \epsilon {\bf E} \otimes {\bf E} + \mu {\bf H} \otimes {\bf H}\,,
\ee
where $\epsilon_\pm = \epsilon_0 \sqrt{c/c_\pm}$ and $\mu_\pm = \mu_0$ are the dielectric and magnetic susceptibilities of each fluid, respectively, $\mu_0$ and $\epsilon_0$ being the ones of vacuum.
From the panels (g) and (h) of Figs.~\ref{fig2} and \ref{fig3}, the comparison of $\Pi_{\rm R}$ and $\Pi_{\rm GC}$~\cite{remark_testEq1} allows to draw the same conclusions as in the acoustic case, including the possible role of a light induced on-axis flow ascribed to light scattering by refractive index critical fluctuations (``optical streaming")~\cite{wunenburger_11, chraibi_11}. As shown in Fig.~\ref{fig2}(e), where $\Pi_{\rm R}$ is plotted for both TE and TM cases, the radiation pressure is almost insensitive to the polarization state of the wave, as expected.

In previous works we demonstrated that the waveguiding of the field along the deformed interface constitutes the feedback mechanism for radiation pressure effects. In particular, this can lead to the stabilization of translationally invariant (here along the $z$ axis) liquid columns by acoustic~\cite{bertin_10} or electromagnetic~\cite{brasselet_08} guided waves. Next, we show that this mechanism applies to noninvariant fluid deformations as well.
We note that OM2 and AM2 deformations have been already observed, however not explained,  in~\cite{casner_06} and~\cite{issenmann_08}, respectively.

\begin{figure*}[t]
\centering\includegraphics[width=1.9\columnwidth]{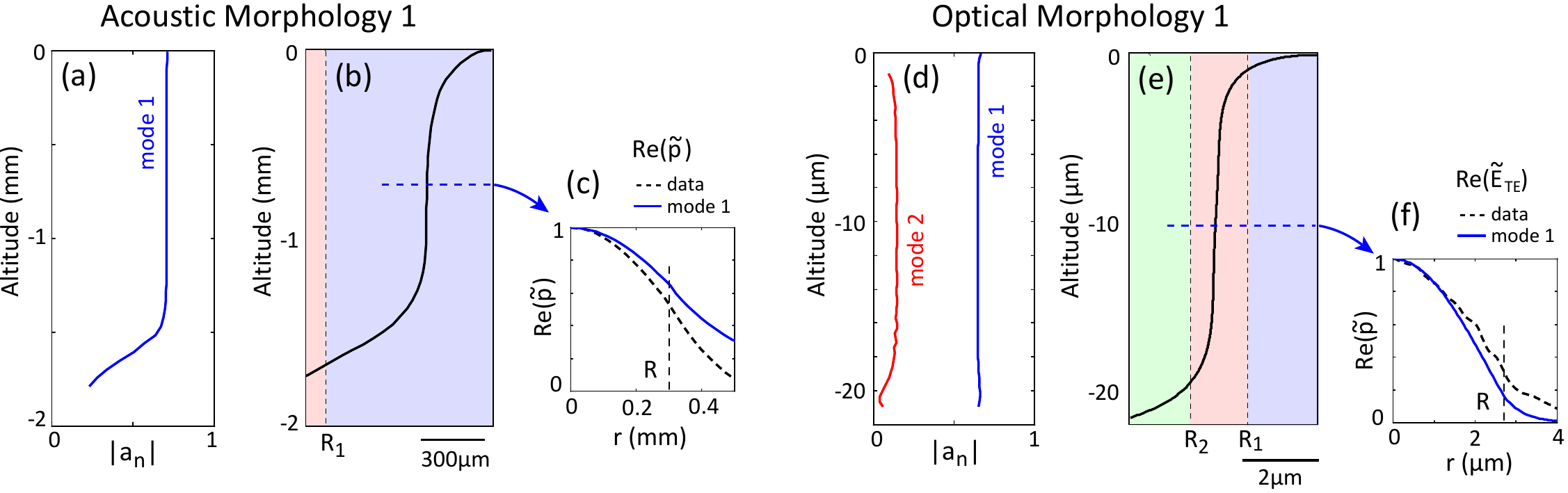}
\caption{\label{fig5}
(color online) Analysis of the AM1 and OM1 cases, see Fig.~\ref{fig1}(e) and Fig.~\ref{fig1}(c). (a,d) $z$-dependence of the modulus of the normalized amplitude $|a_n|$ of the $n$th guided mode. (b,e) Deformed interface profiles. The altitude at the tip of the deformation being set to zero in all cases. Each colored areas refers to radial range of existence of a given set of guided modes. (c,f) Correspondence between the numerically computed field at altitudes that correspond to locally cylindrical portion of a given deformation (solid curve) with the field of the dominant guided mode existing for a cylindrical waveguide with same radius (dashed curve), with $\widetilde \psi(r,z,t) = \frac{\psi(r,z,t) \exp[-i\arg(\psi(0,z,t))]}{\max_r|\psi(r,z,t) |}$ where $\psi = p$ in acoustics and $\psi=E_{\rm TE}$ in optics.}
\end{figure*}

\begin{figure*}[t]
\centering\includegraphics[width=1.9\columnwidth]{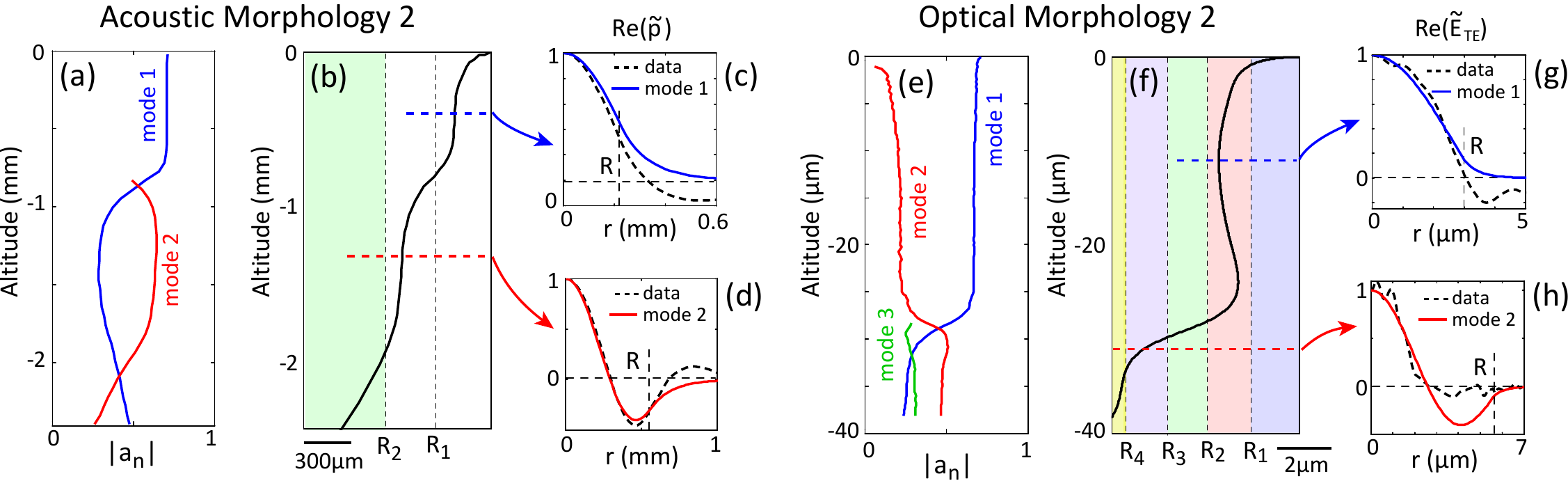}
\caption{\label{fig6}
(color online) As in Fig.~\ref{fig5} for the AM2 and OM2 cases, see Fig.~\ref{fig1}(f) and Fig.~\ref{fig1}(d).}
\end{figure*}

\begin{figure}[t]
\centering\includegraphics[width=0.9\columnwidth]{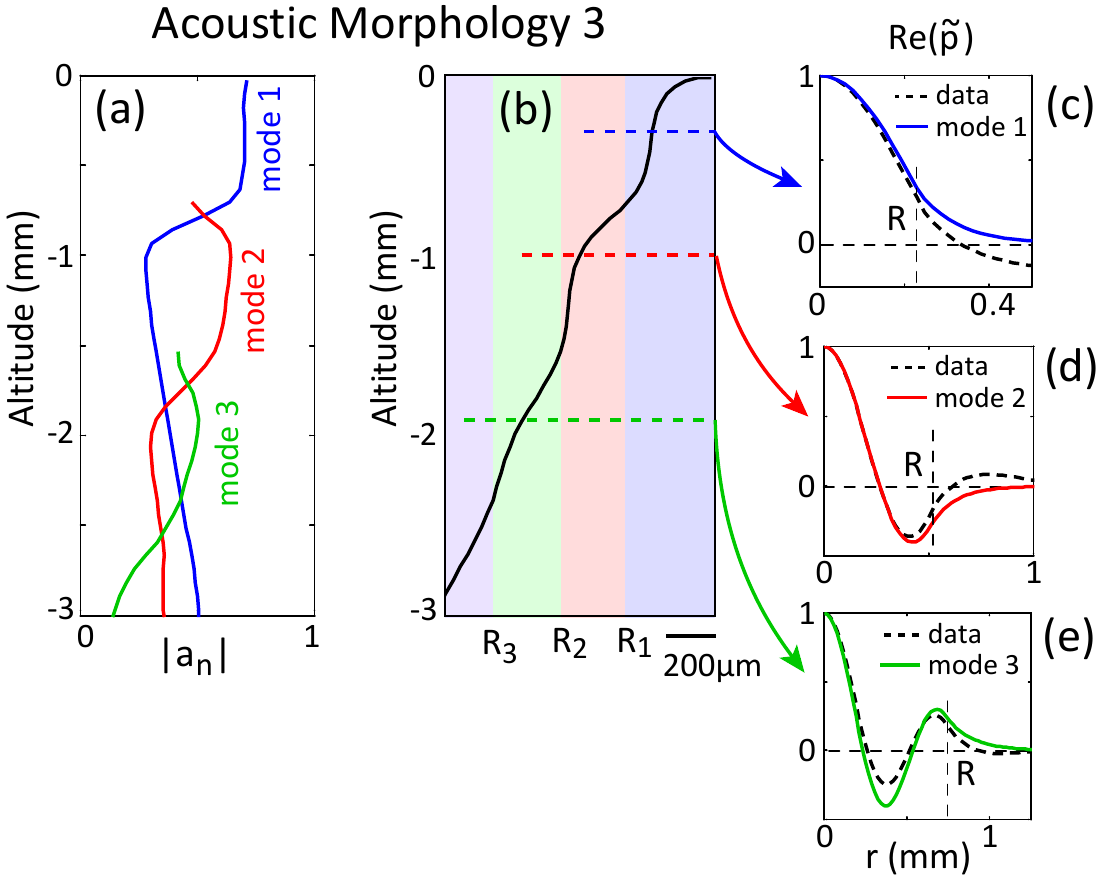}
\caption{\label{fig7}
(color online) As in Fig.~\ref{fig5} for the AM3 case, see Fig.~\ref{fig1}(g).}
\end{figure}

For this purpose, we compare (i) the numerically computed field at altitudes that correspond to locally cylindrical shape of a given deformation with radius $R$ with (ii) the field of the dominant guided mode existing for a cylindrical waveguide with same radius. As shown in Figs.~\ref{fig5},  \ref{fig6} and \ref{fig7}, we observe a striking coincidence that leads us to calculate the modal content of the field propagating along the deformation. The $z$-dependence of the modulus of the normalized amplitude $|a_n|$ of the $n$th guided mode~\cite{remark_an} is shown in Figs.~\ref{fig5}(a) and \ref{fig5}(d), \ref{fig6}(a) and \ref{fig6}(e), and \ref{fig7}(a) for the AM1, AM2, AM3, OM1 and OM2 situations. Note that the fundamental mode $n=1$ of a cylindrical waveguide exists whatever its radius whereas higher order modes $n>1$ exist above cut-off radii $r=R_{n-1}$, respectively. The observed plateaus of $|a_n|$ versus $z$ coincide with locally $z$-invariant morphologies. This indicates that every cylindrical portion of the fluid deformation actually behaves as a waveguide.

Interestingly, the step-like morphologies can be classified using a dimensionless quantity that only depends on the wavelength, the beam waist and the phase velocities of each fluids. Following standard waveguiding theory~\cite{book_waveguiding}, we introduce the characteristic normalized frequency $V_{\rm c} = k_+w_0\sqrt{(c_+/c_-)^2 - 1}$, where $k_+$ is the wavevector in fluid ``$+$". The number of existing guided modes is known to increase with $V_{\rm c}$, so does the order of the observed morphologies. Indeed  $V_{\rm c} = 5.0$, 6.0, 8.1, and 9.0 for the acoustic morphologies AM1, AM2, AM3 and AM$n$ shown in Fig.~\ref{fig1}(e)--(h), respectively, whereas $V_{\rm c} = 3.6$ and 6.1 for the optical morphologies OM1 and OM2 shown in Fig.~\ref{fig1}(c) and \ref{fig1}(d), respectively. However, we notice that a given morphology guides a larger number of modes in optics than in acoustics. This can be grasped by comparing, on the one hand, the cases AM1 and OM1, and, on the other hand, the cases AM2 and OM2. This could be ascribed to different values of $kw_0$, which is a constant of the order of 1 in acoustics whereas it is typically one order of magnitude larger in optics.

We conclude to the observation of universal step-like morphologies of fluid interfaces deformed by the radiation pressure of acoustic and electromagnetic waves. This phenomenon basically relies on the waveguiding properties of sound or light in axisymmetric  two-phase liquid deformations with distinct phase velocities. %A next challenge is the quantitative prediction of such morphologies from ab initio by considering a given acoustic or optical beam impinging on a flat interface at rest, which will involve the dynamical coupling between wave propagation and fluid interface deformation.

%\begin{acknowledgements}
%The authors thank R. Boisgard for useful discussions on acoustic numerical simulations and J. Petit for experimental assistance on the optical set-up.
%\end{acknowledgements}

{\small * Corresponding authors' e-mails : e.brasselet@loma.u-bordeaux1.fr and r.wunenburger@loma.u-bordeaux1.fr}
\vspace{-4mm}
%\bibliography{biblio}

\begin{thebibliography}{0}%
\makeatletter
\providecommand \@ifxundefined [1]{%
 \@ifx{#1\undefined}
}%
\providecommand \@ifnum [1]{%
 \ifnum #1\expandafter \@firstoftwo
 \else \expandafter \@secondoftwo
 \fi
}%
\providecommand \@ifx [1]{%
 \ifx #1\expandafter \@firstoftwo
 \else \expandafter \@secondoftwo
 \fi
}%
\providecommand \natexlab [1]{#1}%
\providecommand \enquote  [1]{``#1''}%
\providecommand \bibnamefont  [1]{#1}%
\providecommand \bibfnamefont [1]{#1}%
\providecommand \citenamefont [1]{#1}%
\providecommand \href@noop [0]{\@secondoftwo}%
\providecommand \href [0]{\begingroup \@sanitize@url \@href}%
\providecommand \@href[1]{\@@startlink{#1}\@@href}%
\providecommand \@@href[1]{\endgroup#1\@@endlink}%
\providecommand \@sanitize@url [0]{\catcode `\\12\catcode `\$12\catcode
  `\&12\catcode `\#12\catcode `\^12\catcode `\_12\catcode `\%12\relax}%
\providecommand \@@startlink[1]{}%
\providecommand \@@endlink[0]{}%
\providecommand \url  [0]{\begingroup\@sanitize@url \@url }%
\providecommand \@url [1]{\endgroup\@href {#1}{\urlprefix }}%
\providecommand \urlprefix  [0]{URL }%
\providecommand \Eprint [0]{\href }%
\providecommand \doibase [0]{http://dx.doi.org/}%
\providecommand \selectlanguage [0]{\@gobble}%
\providecommand \bibinfo  [0]{\@secondoftwo}%
\providecommand \bibfield  [0]{\@secondoftwo}%
\providecommand \translation [1]{[#1]}%
\providecommand \BibitemOpen [0]{}%
\providecommand \bibitemStop [0]{}%
\providecommand \bibitemNoStop [0]{.\EOS\space}%
\providecommand \EOS [0]{\spacefactor3000\relax}%
\providecommand \BibitemShut  [1]{\csname bibitem#1\endcsname}%
\let\auto@bib@innerbib\@empty
%</preamble>
\end{thebibliography}%


\begin{thebibliography}{}
\vspace{-4mm}
\bibitem{loudon_12}
R. Loudon and C. Baxter, Proc. R. Soc. A {\bf 468}, 1825 (2012).

\bibitem{beyer_78}
R. T. Beyer, J. Acoust. Soc. Am. {\bf 63}, 1025 (1978).

\bibitem{hertz_39}
G. Hertz and H. Mende, Z. Phys. {\bf 114}, 354 (1939).

\bibitem{ashkin_73}
A. Ashkin and J. M. Dziedzic, Phys. Rev. Lett. {\bf 30}, 139 (1973).

\bibitem{ostrovskaya_88}
G. V. Ostrovskaya, Sov. Phys. Tech. Phys. {\bf 33}, 468 (1988).

\bibitem{khuri_88}
B. T. Khuri-Yakub {\it et al.}
%, P. A. Reinholdtsen, C-H. Chou, J. F. Vesecky, and C. C. Teague
, Appl. Phys. Lett. {\bf 52}, 1571 (1988).

\bibitem{zhang_88}
J. Z. Zhang and R. K. Chang, Opt. Lett. {\bf 13}, 916 (1988).

\bibitem{elrod_89}
S. A. Elrod {\it et al.}
%, B. Hadimioglu, B. T. Khuri-Yakub, E. G. Rawson, E. Richley, C. F. Quate, N. N. Mansour, and T. S. Lundgren
, J. Appl. Phys. {\bf 65}, 3441 (1989).

\bibitem{guck_00}
J. Guck {\it et al.}
%, R. Ananthakrishnan, T. J. Moon, C. C. Cunningham, and J. K\"as
, Phys. Rev. Lett. {\bf 84}, 5451 (2000).

\bibitem{fatemi_98}
M. Fatemi and J. F. Greenleaf, Science {\bf 280} 82 (1998).

\bibitem{remark_waist}
In the optical case $w_0$ is defined as the Gaussian beam radius at $\exp(-2)$ of its maximal intensity in its focal plane, with $w_0=1.75~\mu$m and $w_0=3~\mu$m for the OM1 and OM2 situations. In the acoustic case, $w_0$ is defined as the waist of a Gaussian fit of the actual intensity profile in the focal plane of the spherical transducer, which gives $w_0 \simeq 0.86 \lambda$ where $\lambda$ is the wavelength [B. Issenmann {\it et al.}, Phys. Rev. Lett. {\bf 97}, 074502 (2006)].

\bibitem{casner_01}
A. Casner and J.-P. Delville, Phys. Rev. Lett. {\bf 87}, 054503 (2001).

\bibitem{delville_05}
J. P. Delville, A. Casner, R. Wunenburger, and I. Brevik, in {\it Trends in Lasers and Electro-Optics Research}, edited by W. T. Arkin (Nova Science, Hauppauge, NY, 2006).

\bibitem{kirkup_98}
S. Kirkup, {\it The Boundary Element Methods in Acoustics}, (Integrated Sound Software, 1998).

\bibitem{remark_testEq1}
In practice, the radial profiles of $\Pi_{\rm R}$ and $\Pi_{\rm GC}$ are normalized by the maximum value of $\Pi_{\rm R}$. Then, $\Pi_{\rm GC}$ is multiplied by a factor $a$ so that $\Pi_{\rm GC}$ fits with $\Pi_{\rm R}$ for $r$ larger than the first maximum of $\Pi_{\rm GC}$. We indeed discard the area close to the $z$ axis where wave-induced flows are suspected to modify Eq.~(1), as explained later in the text. For all measurements, $|a-1|$ is less than the relative uncertainties on the values of $\Pi_{\rm GC}$, which are estimated to be typically 15\% in the acoustic case and 20\% in the optical case following the method presented in [B. Issenmann {\it et al.}, Europhys. Lett. {\bf 83}, 34002 (2008)] and the measurements presented in [R. Wunenburger {\it et al.}, Phys. Rev. E {\bf 73}, 036315 (2006)], respectively. More precisely, in Fig.~\ref{fig2} we have $\Pi_{\rm GC}^{\rm AM1}(r=0)=490\pm70$~Pa and $\Pi_{\rm GC}^{\rm OM1}(r=0)=0.42\pm0.08$~Pa, in Fig.~\ref{fig3} we have $\Pi_{\rm GC}^{\rm AM2}(r=0)=400\pm60$~Pa and $\Pi_{\rm GC}^{\rm OM2}(r=0)=0.42\pm0.08$~Pa, and in Fig.~\ref{fig4} we have $\Pi_{\rm GC}^{\rm AM3}(r=0)=250\pm40$~Pa. In addition, we note that the latter uncertainties mainly depend on the accuracy of the measurement of the interfacial tension.

\bibitem{book_acousticstreaming}
{\it Nonlinear Acoustics}, edited by M. F. Hamilton and D. T. Blackstock (Academic, New York, 1997).

\bibitem{ikelle_12}
L. T. Ikelle, Geophys. J. Int. {\bf 189}, 1771 (2012).

\bibitem{remark_electrostriction}
The electrostriction term is discarded in Eq.~(\ref{eq_Tem}) since it does not contribute to the stress balance at the interface [H. Chraibi {\it et al.}, Eur. J. Mech. B/Fluids {\bf 27}, 419 (2008)].

\bibitem{wunenburger_11}
R. Wunenburger {\it et al.}
%, B. Issenmann, E. Brasselet, C. Loussert, V. Hourtane, and J.-P. Delville
, J. Fluid Mech. {\bf 666}, 273 (2011).

\bibitem{chraibi_11}
H. Chra\"ibi {\it et al.}
%, R. Wunenburger, D. Lasseux, J. Petit, and J.-P. Delville
, J. Fluid Mech. {\bf 688}, 195 (2011).

\bibitem{bertin_10}
N. Bertin, R. Wunenburger, E. Brasselet, and J.-P. Delville, Phys. Rev. Lett. {\bf 105}, 164501 (2010).

\bibitem{brasselet_08}
E. Brasselet, R. Wunenburger, and J.-P. Delville, Phys. Rev. Lett. {\bf 101}, 014501 (2008).

\bibitem{casner_06}
A. Casner, J.-P. Delville, and I. Brevik, J. Opt. Soc. Am. B {\bf 20}, 2355 (2003).

\bibitem{issenmann_08}
B. Issenmann {\it et al.}
%, A. Nicolas, R. Wunenburger, S. Manneville, and J.-P. Delville
, Europhys. Lett. {\bf 83}, 34002 (2008).

\bibitem{remark_an}
We define $a_n = \langle \psi \psi_n\rangle/\sqrt{ \langle \psi \psi\rangle \langle \psi_n \psi_n\rangle}$ where $\psi \equiv p$ within the acoustic framework, $n$ refers to the $n$th guided mode, and $\langle \varphi_a \varphi_b \rangle = \frac{\pi}{i\omega\rho}\int_0^\infty(\varphi_a \frac{\partial \varphi_b^*}{\partial z} + \varphi_b^* \frac{\partial \varphi_a}{\partial z})rdr$ where asterix indicates complex conjugation. In practice, the upper bound of the latter integral is taken as the local radius $R(z)$ of the deformation at altitude $z$.

\bibitem{book_waveguiding}
K. Okamoto, {\it Fundamentals of Optical Waveguides}, 2nd ed. (Elsevier, Amsterdam, 2006).



\end{thebibliography}
\bibliographystyle{apsrev4-1}

%\newpage

%---------------------------------------------------------------------------------------------------
%---------------------------------------------------------------------------------------------------

%---------------------------------------------------------------------------------------------------
%---------------------------------------------------------------------------------------------------

%---------------------------------------------------------------------------------------------------
%---------------------------------------------------------------------------------------------------

%---------------------------------------------------------------------------------------------------
%---------------------------------------------------------------------------------------------------

%---------------------------------------------------------------------------------------------------
%---------------------------------------------------------------------------------------------------

%---------------------------------------------------------------------------------------------------
%---------------------------------------------------------------------------------------------------

\end{document}